\documentclass[twocolumn,nofootinbib,amsmath,prd,aps,superscriptaddress,tightenlines,preprintnumbers]{revtex4}
\usepackage{graphicx}
\usepackage{epstopdf}
\usepackage{amsfonts}
\usepackage{bm}
\usepackage{epsfig}
\usepackage{graphics}
\usepackage{xspace}
\usepackage[usenames]{color}
\def\bea{\begin{eqnarray}}
\def\eea{\end{eqnarray}}
\def\ba{\begin{eqnarray}}
\def\ea{\end{eqnarray}}
\def\be{\begin{equation}}
\def\ee{\end{equation}}
\def\beq{\begin{equation}}
\def\eeq{\end{equation}}
\newcommand{\slashed}{\slash \hspace{-0.23cm}}
\newcommand{\lsim}{\mathrel{\rlap{\lower4pt\hbox{\hskip1pt$\sim$}}
    \raise1pt\hbox{$<$}}}         
\newcommand{\gsim}{\mathrel{\rlap{\lower4pt\hbox{\hskip1pt$\sim$}}
    \raise1pt\hbox{$>$}}}         

\def\nn{\nonumber}

\begin{document}

\newcount\hour \newcount\minute
\hour=\time \divide \hour by 60
\minute=\time
\count99=\hour \multiply \count99 by -60 \advance \minute by \count99
\newcommand{\mydate}{\ \today \ - \number\hour :00}

\title{Forward-backward asymmetry in $t\bar t$ production from flavour symmetries}

\author{Benjam\'\i{}n Grinstein}
\email[Electronic address:]{bgrinstein@physics.ucsd.edu}
\affiliation{Department of Physics, University of California at San Diego, La Jolla, CA 92093}

\def\Cincy{Department of Physics, University of Cincinnati, Cincinnati, Ohio 45221,USA}

\author{Alexander L. Kagan}
\email[Electronic address:]{kaganalexander@gmail.com} 
\affiliation{\Cincy}
\affiliation{ Department of Particle Physics \& Astrophysics, Weizmann Institute of Science, Rehovot 76100, Israel}

\author{Michael Trott}
\email[Electronic address:]{mtrott@perimeterinstitute.ca} 
\affiliation{Perimeter Institute for Theoretical Physics, Waterloo, ON N2J-2W9, Canada}

\author{Jure Zupan}
\email[Electronic address:]{jure.zupan@cern.ch}
\email[On leave of absence from Josef Stefan Institute and
U. of Ljubljana, Ljubljana, Slovenia.]{}
\affiliation{\Cincy}

\begin{abstract}
We show that the forward-backward asymmetry in top quark pair production 
can be enhanced by fields that transform nontrivially under the flavour group and satisfy Minimal Flavour Violation, while at the same time the constraints from associated effects on the $d \sigma(t \, \bar{t})/d M_{t \bar{t}}$ distribution, dijet resonance searches, same sign top pair production and other phenomenology are satisfied. We work out two examples in detail, one where a scalar colour anti-sextet field, that is also anti-sextet of $\rm SU(3)_U$, enhances the forward-backward asymmetry, and one where the enhancement arises from a vector colour octet field that is also an octet of $\rm SU(3)_U$. 
\end{abstract}

\maketitle
\newpage
{\bf Introduction.} In the Standard Model (SM) the only quark that couples with $O(1)$ strength to the electroweak symmetry breaking sector -- the SM Higgs boson -- is the top quark. 
Anomalous interactions of the top quark could thus be our first window on the physics that stabilizes the electroweak scale. The large coupling of the top to the Higgs
also breaks the SM quark flavour group $\rm G_F=SU(3)_U\times SU(3)_D\times SU(3)_Q$, which arises in the limit where the SM Yukawas ($Y_U,Y_D$) vanish,
\begin{equation}
{\cal L}_Y=(Y_{U})^i_j \,{\bar u}_{iR} \, Q_L^j \, H+(Y_{D})^i_j \, {\bar d}_{iR} \, Q_L^j \, H ^{\dagger}+{\rm h.c.}. 
\end{equation}
It is reasonable to expect that New Physics (NP) which stabilizes the electroweak scale could have non trivial flavour structure, and that measurements of top quark properties could also improve our understanding
of the origin of flavour.

Recently, CDF announced that for the mass of the $t\bar t$ pair $M_{\bar{t} \, t} \geq 450 \,{\rm GeV}$, the measured asymmetry 
in top quark pair production, $\rm A_{FB}^{t \bar{t}}=0.475\pm0.114$  \cite{Aaltonen:2011kc}, differs by $3.4\sigma$ from the next-to-leading order (NLO) SM  prediction $\rm A_{FB}^{t \bar{t}}=0.088\pm0.013$.
This reinforces anomalously large past measurements of $\rm A_{FB}^{t \bar{t}}$
by CDF and $\rm DO \! \! \! \!/$ \cite{Aaltonen:2008hc,:2007qb,DOconfNote}.

This discrepancy could be due to an additional NP contribution to $t\bar t$ production or a statistical fluctuation. 
NP models that explain these anomalies generally have new particles exchanged in the $s$ or $t$ channel, and each case faces a number of challenges. For models with $s$ channel exchange, there is no evidence for a resonance in the $t\bar t$ invariant mass spectrum \cite{Aaltonen:2009iz,Aaltonen:2009tx}, pushing the mass of the particle to at least $\mathcal{O}(1{\rm TeV})$ \cite{Cao:2010zb}. Further, in order to obtain a positive 
$\rm A_{FB}^{t\bar t}$, the NP vector field has to couple to light quarks with an opposite sign than to the top \cite{Ferrario:2009bz,Cao:2010zb,Frampton:2009rk}. These couplings can lead to large flavour changing neutral currents (FCNC's), violating the observed agreement between the SM and FCNC observables.  In \cite{Delaunay:2011vv,Bai:2011ed} $\rm SU(2)_U$ symmetric quark couplings suppress dangerous $\rm D$ meson mixing. 

Models with $t$ channel exchange require large inter-generational couplings.
It is possible to arrange that despite such couplings no observable FCNC effects arise \cite{Shelton:2011hq,Jung:2009jz,Dorsner:2010cu}; however, it is challenging for such scenarios to be incorporated into a realistic model of flavour. In concrete models, nonzero $\bar c -u$ couplings can lead to unacceptably large contributions to $\rm D$ meson mixing.
$t$ channel exchanges can also lead to same sign top-pair production, which is tightly constrained \cite{Aaltonen:2008hx}.

This letter demonstrates that all of the above obstacles, as well as dijet constraints, can be overcome, if the NP particle exchanged in the $t$ or $s$ channel is in a nontrivial representation of  $\rm G_F$. For instance, 
an $s$ channel exchange of an $\rm SU(3)_U$ octet vector field automatically has couplings to light quarks of opposite sign than to the top quark
\beq
\begin{split}
 \tfrac{1}{\sqrt3}V_\mu^8\big(\bar u_R \gamma^\mu u_R+\bar c_R \gamma^\mu c_R-&2\bar t_R \gamma^\mu t_R) +\cdots,
\end{split}
\eeq
where the ellipses denote the remaining field components of the $\bf 8$ representation. The same flavour octet vector will also lead to $u\leftrightarrow t$ transitions in the $t$ channel
\beq
\begin{split}
(\bar U_R T^A\gamma^\mu U_R)V_\mu^A=&\big(V_\mu^4-i V_\mu^5\big)\big(\bar t_R \gamma^\mu u_R\big)+\cdots,
\end{split}
\eeq
from the exchange of $V_\mu^{4,5}$ octet components, and similarly to $c\leftrightarrow t$ transitions from the exchange of $V_\mu^{6,7}$. Despite the large intergenerational transitions, no FCNC's arise before flavour breaking. Integrating out the NP, one matches onto 4-quark operators 
that are schematically of the form $(\bar q_i q_j) \Delta_{ij,kl} (\bar q_k q_l)$. 
Before flavour breaking, $\Delta_{ij,kl}$ has a form 
$\Delta_{ij,kl}=\dots \delta_{ij,kl}+\dots \delta_{il}\delta_{jk}$. 
One generates four quark operators $(\bar u c)(\bar c u)$ but not $(\bar u c)^2$.

For $\rm G_F$ breaking we assume 
Minimal Flavour Violation (MFV), where the flavour violation (FV) in the NP sector is proportional to $Y_U,Y_D$  \cite{Chivukula:1987py,Hall:1990ac,D'Ambrosio:2002ex}. As a result FCNC's are consistent with experiment even for NP mass scales $< {\rm TeV}$. We use the MFV formalism of \cite{D'Ambrosio:2002ex}.


{\bf General analysis.}
First, we study in a model independent way if NP contributions need to interfere with the SM in order to obtain the observed $\rm A_{FB}^{t \bar{t}}$. Let $\sigma_{F,B}^{SM}$ and $\sigma_{F,B}^{NP}$ be the SM and NP forward and backward cross sections, respectively \cite{Cao:2010zb}.  The latter contain the contributions from NP interfering with the SM and from the NP-matrix elements squared. If interference dominates, $\sigma_{F,B}^{NP}$ can have either sign, if interference is negligible, these terms have to be positive. 

\begin{figure}
\includegraphics[width=0.40\textwidth]{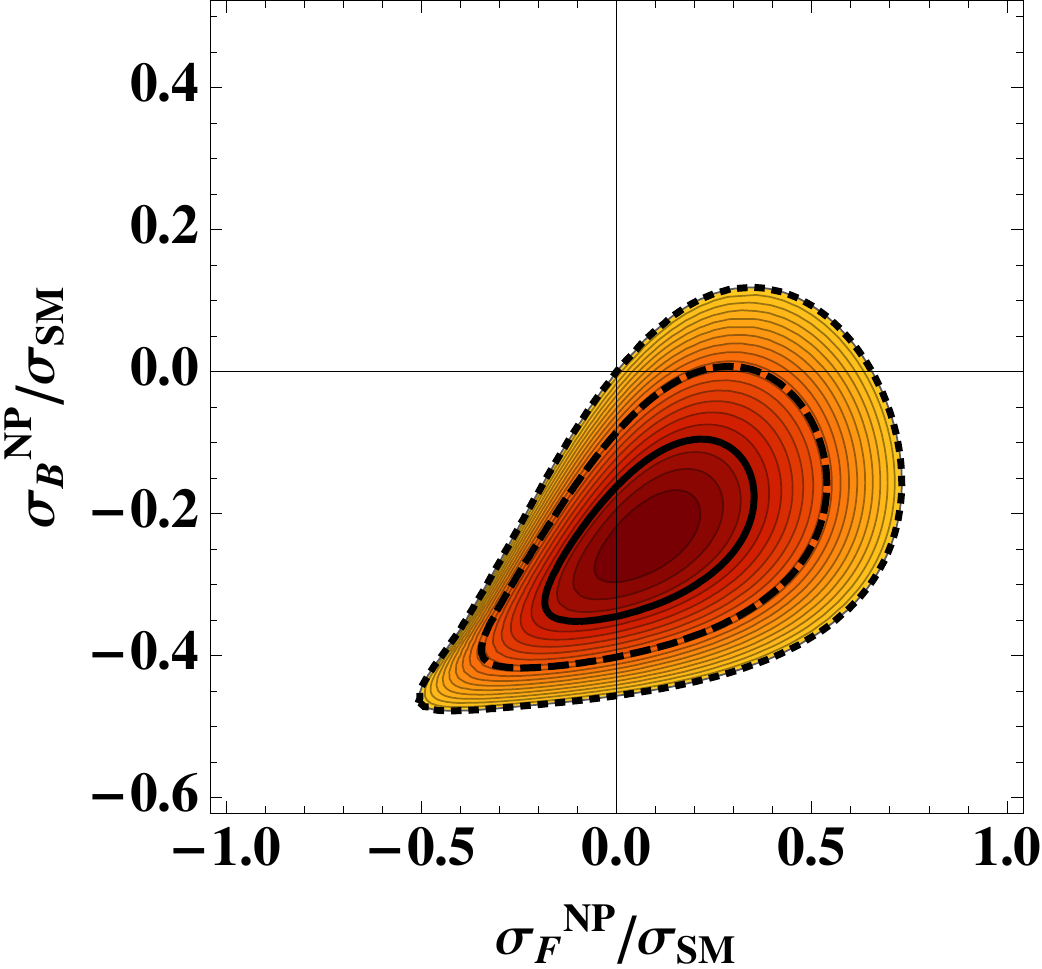}
\caption{$\sigma_F^{NP}$ and $\sigma_B^{NP}$ (normalized to the SM values) needed to explain the measured $\rm A_{FB}^{t \bar{t}}$, while at the same time being compatible with ${\sigma}(t \, \bar{t})$. The contours correspond to 1-$\sigma$ (solid), 2-$\sigma$ (dashed) and 3-$\sigma$ (dotted) allowed regions. 
}
\label{f.sigmas}
\end{figure}

We use the measured and predicted values of  $\rm A_{FB}^{t\bar t}=0.475\pm0.114$,
and $0.088\pm0.013$ respectively, for $M_{t\bar t}>450 {\rm \,GeV}$,
together with the 
measured total cross section $\sigma(t\bar t)(M_{t \bar t}> 450 {\rm \,GeV})=1.9\pm 0.5~{\rm pb}$  \cite{Aaltonen:2009iz}
and the predicted value $ \sigma^{SM}(M_{t\bar t}>450 {\rm GeV})=2.26\pm 0.18~{\rm pb}$ (using \cite{Ahrens:2010zv}) to derive the constraints on $\sigma_F^{NP}$, $\sigma_B^{NP}$ shown in Fig. \ref{f.sigmas}.  
We find a preference for $\sigma_F^{NP} >0$ and $\sigma_B^{NP} <0$ at two $\sigma$ significance, which points to an interference effect. If the $s$ channel contribution dominates, this means that the exchanged particle has to be a colour octet vector. The other options are large $t$ channel interference, or a combination of both channels.
 


{\bf Models.} There are 22 (14) possible quantum number assignments for vector (scalar) fields that couple to quark bi-linears through marginal interactions  and conserve  $\rm G_F$ \cite{Arnold:2009ay,LongPaper} without the insertion of $Y_U,Y_D$. In this letter, we discuss two of the models that are consistent with the general analysis. The first model contains a scalar field, $S$, transforming as $(\bar{6},1)_{-4/3}$ under the SM gauge group $\rm SU(3) \times SU_L(2) \times U_Y(1)$ \cite{Arnold:2009ay} and as $(\bar 6,1,1)$ under $\rm G_F$. The second model contains a vector field, $V$, that is an $(8,1)_{0}$ under the SM gauge group, and $(8,1,1)$ under $\rm G_F$.

{\underline {\sl Anti-Sextet scalar.}} The scalar fields are in the anti-sextet of colour and the anti-sextet of $\rm SU(3)_U$ such that $S^{\alpha\beta}_{kl}=S^{\beta\alpha}_{kl}=S^{\alpha\beta}_{lk}$ where $\alpha, \beta$ are colour indices and generation indices are $k,l$. Consider the $\rm G_F$ symmetric Lagrangian
\begin{equation}
\mathcal{L}_{S} = \eta_1 \, U_{R \alpha}^k \, U_{R_\beta}^l \, S_{kl}^{\alpha \, \beta} + h.c.,
\end{equation}
with $U_R=(u_R,c_R,t_R)$. The large top Yukawa $y_t= \sqrt{2} m_t/v$ breaks $\rm SU(3)_U$ so that ${\cal L}_S$ gets corrections for couplings involving the top quark. In MFV, the corrections are 
\begin{equation}
\Delta \mathcal{L}_{S} =\eta_2 \, U_{R \alpha}^k \, [\Delta_U U_{R_\beta}]^l \, S_{kl}^{\alpha \, \beta} + h.c.,
\end{equation}
with additional corrections from higher insertions of $\Delta_U=Y_U Y_U^\dagger$. The couplings of $S$ to the first two generations are given by $\eta_{ij}=\eta_1$, while the $t-u$ and $t-c$ couplings
are $\eta_{i3}=\eta_1+y_t^2 \eta_2$. (Here and below $i,j$ run over $1,2$ while $k,l$ run over $1,2,3$.)
The FV also splits the masses of $S_{kl}$,  so that $m_{S_{3k}}^2- m_{S_{ij}}^2 \propto y_t^2 m_{S_{ij}}^2$.


{\underline {\sl Octet vector}}. For the vector field we use the notation $V\equiv V_{A,B} ({\cal T}^A)_\alpha^\beta (T^B)_k^l /4$, with ${\cal T}^A$ and $T^B$ the colour and flavour $\rm SU(3)$ matricies with normalization ${\rm Tr}[T^A T^A] = 2$ . The flavour symmetric interaction Lagrangian is then
\beq
\mathcal{L}_{V} =  \eta_1 \, \bar{U}_R \, \slashed \, V U_R,
\eeq
where $\eta_1$ is real.  The FV corrections from $y_t$ are 
\beq
\Delta \mathcal{L}_{V} =  [\eta_2 \, \bar{U}_R \, \slashed \, V \, \Delta_U \, U_R  +h.c.]+ \eta_3 \, \bar{U}_R \Delta_U\, \slashed \, V \, \Delta_U \, U_R,
\eeq
up to higher insertions of $\Delta_U$. The couplings of the first two generation quarks with the vector are given by $\eta_{ij}=\eta_1$, the $\bar u-t$ and $\bar c-t$ couplings to the vector are given by $\eta_{i3}=\eta_1+y_t^2 \eta_2$, while the $\bar t-t$ coupling is given by $ \eta_{33}=\eta_1+2 y_t^2 {\rm Re}\,[\eta_2]+y_t^4 \eta_3$.  Note that $\eta_2$ can be complex.

\begin{figure}[t]
\includegraphics[width=0.47\textwidth]{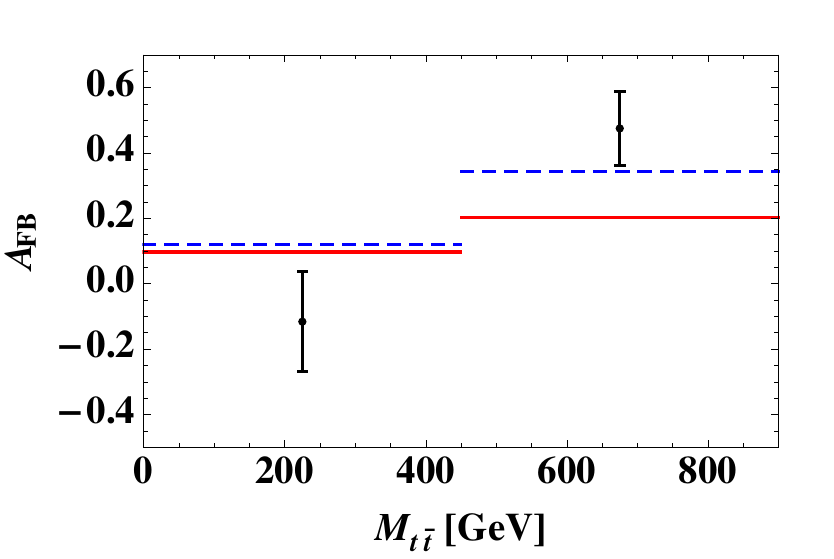}
\includegraphics[width=0.47\textwidth]{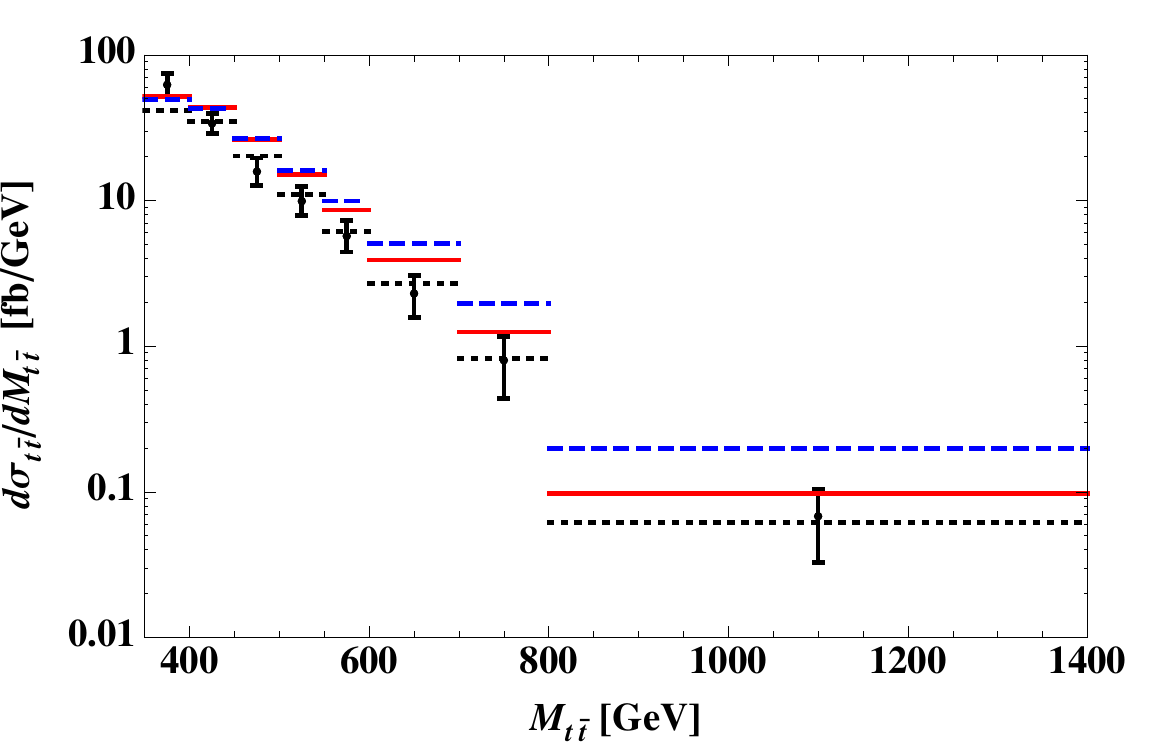}
\caption{$\rm A_{FB}^{t \bar{t}}$ with anti-sextet scalar exchange, compared to experiment (1$\sigma$),
and the corresponding  $d\sigma(t \, \bar{t})/dM_{t\bar t}$ spectrum compared to the measured one.
Also shown is the NLO+NNLL SM spectrum prediction (dotted black). The 
parameter choices are ($m_{S_{k3}} , \eta_{k3}, \Gamma_{S_{k3}}/m_{S_{k3}}$): solid red ($390 \, {\rm GeV}, 0.51, 0.1$); dashed blue ($1300 \, {\rm GeV}, 1.5,  0.5$).
}\label{Fig.scalar}
\end{figure}

\begin{figure}[t]
\includegraphics[width=0.475\textwidth]{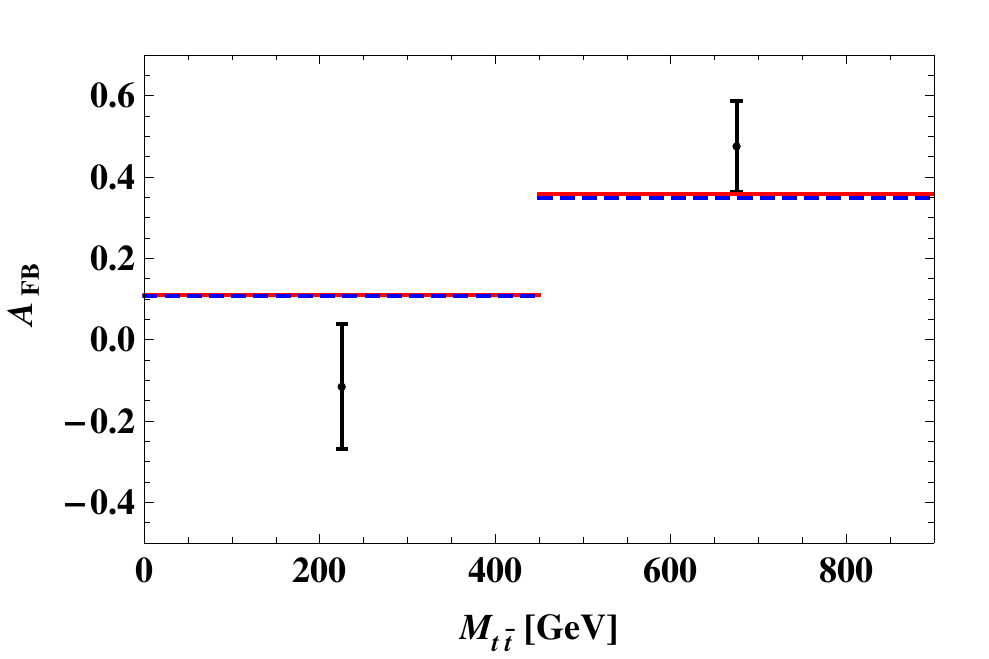}
\includegraphics[width=0.47\textwidth]{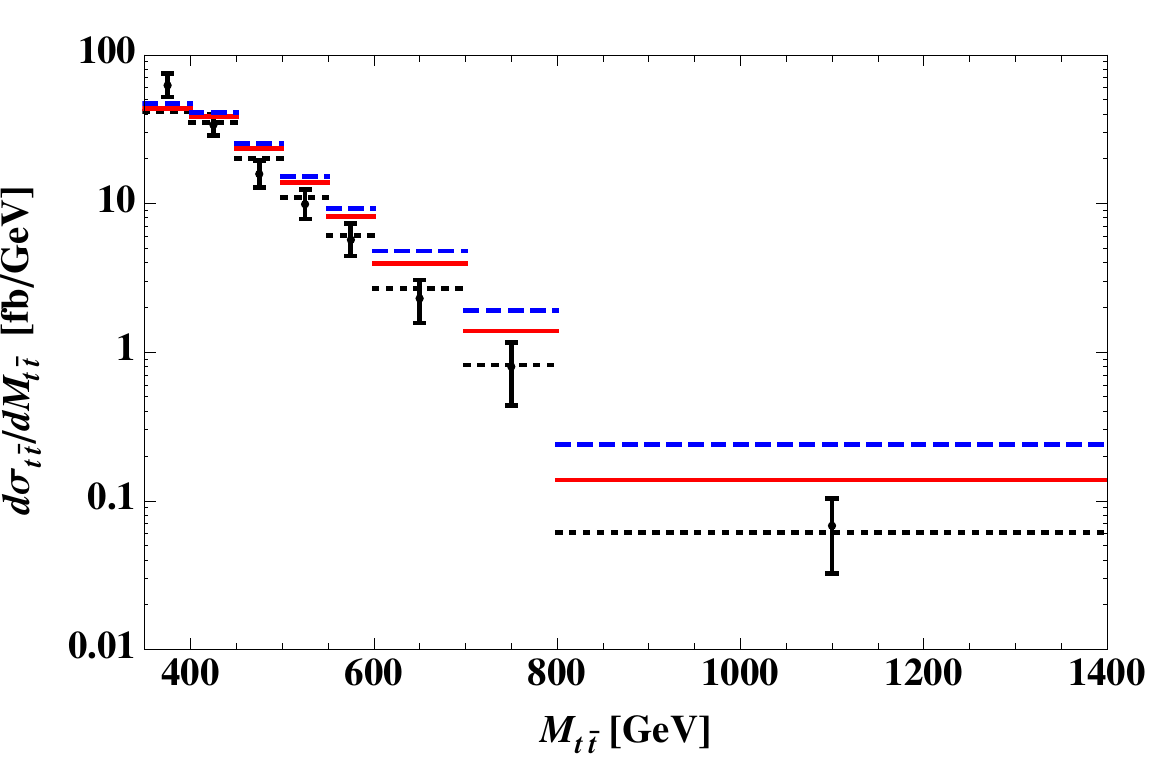}
\caption{$\rm A_{FB}^{t \bar{t}}$ and $d\sigma(t \, \bar{t})/dM_{t\bar t}$ with octet vector exchange, for two 
different values of ($m_{V},  \sqrt{\eta_{ij}\eta_{33}}, \eta_{i3}, \Gamma_V/m_V$): solid red ($300 \, {\rm GeV}, 1,1.33, 0.08$);
dashed blue ($1200 \, {\rm GeV}, 2.2, 4.88, 0.5$), that give approximately the same $\rm A_{FB}^{t \bar{t}}$ in the high mass bin.
%
}\label{Fig.Vector.AFB}
\end{figure}

{\bf Phenomenology.} We use MSTW2008 PDF's \cite{Martin:2009iq} and work to NLO for the SM $t\bar t$ cross section, while we work to LO for the NP corrections to it (including the interference with the SM).
A challenge for any model that seeks to explain the anomalous $\rm A_{FB}^{t \bar{t}}$ measurement is the
agreement of the SM prediction and measurement
for the $d\sigma(t \, \bar{t})/dM_{t\bar t}$ differential cross section  \cite{Aaltonen:2009iz}. The integrated cross section from this measurement is $\hat{\sigma}(t\bar t)=6.9 \pm 1.0$~pb, which should be compared to the NLO prediction with NNLL summation of threshold logarithms, which gives $\sigma(t\bar t)=6.30 \pm 0.19^{+0.31}_{-0.23}$~pb \cite{Ahrens:2010zv}.  The agreement between the SM prediction and this measurement is the most important constraint on our models, and limits the size of $\rm A_{FB}^{t\bar t}$. We collect the predictions for the two models in Figs. \ref{Fig.scalar}, \ref{Fig.Vector.AFB}.
Throughout we use $m_t = 173.1 \, {\rm GeV}$ and use the predictions of \cite{Ahrens:2010zv}. 

%

{\underline {\sl Anti-Sextet scalar.}} In $t\bar t$ production, the flavour anti-sextet scalar can be exchanged in the $t$ channel. The size and the shape of the contribution is controlled by the couplings to the top: $\eta_{k3}=\eta_1+y_t^2 \eta_2$, the mass of the exchanged scalars, $m_{S_{k3}}$, and by their decay widths $\Gamma_{S_{k3}}$. In Fig. \ref{Fig.scalar} we show two representative cases for a light and heavy scalar. A larger value of $\rm A_{FB}^{t\bar t}$ requires a larger deviation from the measured $d \sigma(t \, \bar{t})/d M_{t \bar{t}}$ spectrum, so that agreement within one $\sigma$ of the central value of  $\rm A_{FB}^{t\bar t}$ (for $M_{\bar{t} \, t} \geq 450 \,{\rm GeV}$)
without distorting the shape of the $d\sigma/dM_{t\bar t}$ distribution is challenging. 
Our results qualitatively agree with the (non-MFV, flavour singlet) results of \cite{Shu:2009xf}; the model we discuss is less constrained by $tt$ production constraints and $ \rm D$ mixing phenomenology than \cite{Shu:2009xf} due to its flavour structure. 

Heavier scalar masses require larger values of $\eta_{i3}$ and, compared to lower scalar masses, they distort the $d \sigma/d M_{t \bar{t}}$ distribution more significantly in the high $M_{t \bar{t}}$ region,
where the SM falls steeply in agreement with experiment. However, the PDF uncertainty also rises with large $M_{t\bar{t}}$. 
The contribution to the total cross section from NP is $\sigma(t \, \bar{t})_{NP} \approx (2,2.2)$ pb for the (light, heavy) scalar, which gives $\sigma(t\bar t)_{SM} + \sigma (t \bar t)_{NP}$ 
consistent within 2 $\sigma$ with $\hat{\sigma}(t \, \bar{t})$.
The corresponding integrated asymmetry is $\rm A_{FB}^{t \bar{t}} \approx (0.12, 0.18)$. This result  is
to be compared with the parton level unfolded asymmetry $\rm A_{FB}^{t \bar{t}} = 0.158 \pm 0.075 ({\rm stat,sys})$\cite{Aaltonen:2011kc}.

{\underline {\sl Vector octet.}} In this model the vector contributes to $t \bar t$ production in both the $s$ and
$t$ channels.  
Note that for an $s$ channel exchange, resonance searches in $t \, \bar{t}$ production are relevant \cite{Aaltonen:2009tx} when $m_V > 2 \, m_t$. The couplings we consider are consistent with these searches.

In Fig. \ref{Fig.Vector.AFB} we show examples illustrating a heavy or light vector exchange. For low vector masses, good agreement with both the $\rm A_{FB}^{t\bar t}$ distribution and the $d\sigma/dM_{t\bar t}$ distribution
is possible.
The contribution to the total cross section from NP is $\sigma(t \, \bar{t})_{NP} \approx (1.0, 1.8)$pb 
for the (light, heavy) vectors, consistent with $\hat{\sigma}(t \, \bar{t})$. 
The corresponding inclusive asymmetry is $\rm A_{FB}^{t \bar{t}} \approx (0.17, 0.17)$ for these cases. 

These examples illustrate that for light vector masses the $\eta_{kl}$ can be  
nearly or exactly $\rm G_F$ symmetric, whereas for heavy vectors only moderate $\rm G_F$ breaking due to the top Yukawa insertions is required.


{\underline {\sl Dijet constraints.}} The NP models we consider are initially flavour symmetric in the coupling of the 
scalar and vector fields to the quarks. 
 Dijet resonance searches  \cite{Aaltonen:2008dn} can constrain the couplings to the light quarks for these models.
However, a large flavour breaking due to $y_t$  can lead to the light quarks coupling to the field with $\eta_{ij}=\eta_1$, while the top couples with $\eta_{3j}=\eta_1+y_t^2 \eta_2$ and $\eta_{33}= \eta_1+y_t^2 \eta_2+y_t^4 \eta_3$ (the latter being relevant for the vector) reducing the strength of any constraints on the mass scale. (Even larger hierarchies can come about from higher orders of $y_t$ insertions, which can be studied in the GMFV formalism \cite{Feldmann:2008ja,Kagan:2009bn}).

We have studied the partonic dijet production in these models and compared the resulting dijet invariant mass spectra to \cite{Aaltonen:2008dn}.
For the anti-sextet scalar, we find that the couplings in the examples we have presented are consistent with \cite{Aaltonen:2008dn}.
For the vector field, we find that light quark couplings $\eta_{ij} \lsim 1$ in the low mass case are consistent with \cite{Aaltonen:2008dn}, while
for the large mass case, light quark couplings $\eta_{ij} \lsim 3$ are consistent.
In both the scalar and vector examples, with light quark couplings in these ranges, we find negligible impact on the dijet invariant mass spectra measured at the LHC \cite{Khachatryan:2010jd}.
Dijet angular distribution studies at the Tevatron \cite{:2009mh}
and at the LHC \cite{Collaboration:2010eza,Collaboration:2011as}
also have the potential to place 
bounds on these models.   Again, for the above light quark couplings, we find that our vector examples above are consistent with the former and have negligible impact on the latter.
For angular distributions at LHC and Tevatron, for the heavy scalar example, a splitting is required so that the light quark coupling is $\lsim 0.8$.

Thus, after including dijet constraints, we find that for the light vector and scalar examples 
the nearly $\rm G_F$ symmetric limit is viable. 
In the heavy vector and scalar examples we find that the dijet constraints require a 
modest breaking of $\rm G_F$.

{\underline {\sl Other constraints and collider signals}}. 
Since the fields are in a nontrivial representation of $\rm G_F$ no significant $tt$ production occurs before flavour breaking.
The leading generation of $t \, t$ is suppressed by $\sim (y_c y_t y_b^2 V_{cb})^2$ times the charm quark PDF suppression, making $t \, t$ production negligible. These models are also consistent with low energy flavour violation measurements due to their flavour symmetry structure. The contributions  to $\rm  D-\bar D$ mixing for the two models are $\sim (y_u y_c y_b^2)^2$ suppressed.  Other MFV models with fields that transform nontrivially under $\rm G_F$ and which address $\rm A_{FB}^{t\bar t}$ may have potentially interesting effects in $\rm B_{s,d}$ meson mixing \cite{LongPaper}, along the lines discussed in \cite{Kagan:2009bn,Trott:2010iz}.
 
Our findings suggest that a viable explanation of the CDF measurements of $\rm A_{FB}^{t\bar t}$  
will be accompanied by enhanced $t\bar t$ production at large invariant masses.  
For example, in Figs.~2,3 and for the light vector and scalar examples, there is an excess of roughly a factor of 2
in the $M_{t\bar t} > 800$ GeV bin between the predicted and SM integrated cross sections (with larger excesses for heavier particles).
It is interesting that such a hierarchy is consistent with the recently reported excess of boosted jets in the top mass window 
with $p_T > 400$ GeV \cite{cdfboostedtops} (also see \cite{Eshel:2011vs}).

We have shown that the CDF measurements of $\rm A_{FB}^{t\bar t}$ and $\sigma(t\bar t)$ possibly point toward light NP particles. If this is the case, they can
be produced at the LHC through $q \, \bar{q}$ and $q \, q$ initial states with significant event rates. The production cross section in these cases depends on the flavour conserving coupling $\eta_{ij}$ to the light quarks which can be $O(1)$.  An interesting possibility is single production at LHC through $\bar u_R  \, \slashed \, V t_R $ or $\bar u^c t_R S$ couplings.  $V$'s are then produced in association with $t$ (or $\bar t$ for the scalar) and searches depend on the $V,S$ decay channels. The dominant decays are through flavour universal couplings to light quarks, since decays to two light jets are kinematically favored. The signal at LHC would then be an excess in the $t+2j$ channel with a resonant structure in the dijet mass spectrum.  If the $\eta_{i3}$ couplings dominate due to significant $y_t$  breaking of $\rm G_F$, then $V$ would decay to $\bar t+j$ (or $t+j$ for $S$). In this case the signature would be $t\bar t+j$, which is potentially observable already with $1 {\rm fb}^{-1}$ data at $7$ TeV \cite{Gresham:2011dg}. 
These states can also be produced through higher dimensional operators or loop suppressed interactions. 
We leave a detailed study of the production rates to a future work.
 
{\bf Conclusions.} 
In NP models that can explain the recently reported $\rm A_{FB}^{t \bar{t}}$ measurements \cite{Aaltonen:2011kc} there is significant tension with other experimental measurements: notably the $t\bar t$ differential cross section, same-sign top pair production, dijet constraints and FCNC's. We have shown that in flavour symmetric models, where the fields transform under $\rm G_F$ and flavour breaking is consistent with MFV, this tension can be reduced
in a simple unified framework. Given present data a light colour and flavour octet vector is preferred.


{\bf Acknowledgements:} We would like to thank J. Kamenik, A. Pierce, the Weizmann Institute Phenomenology group, and K. Zurek for useful discussions. B.~G. is supported in part by the US DoE under contract DE-FG03- 97ER40546, A.~K. is supported by DOE grant FG02-84-ER40153. We thank the authors of \cite{Ligeti:2011vt} for communications regarding the incorrect results of \cite{Shu:2009xf}, which
led to incorrect numerics for the scalar $\bf 6$ in the first version of this paper. 


\end{document}